\documentclass[12pt,a4paper,final]{iopart}
\usepackage{iopams}  
\usepackage{graphicx}
\expandafter\let\csname equation*\endcsname\relax
\expandafter\let\csname endequation*\endcsname\relax
\usepackage{amsmath}
\usepackage{hyperref}
\usepackage{subfigure}
\usepackage{cite}
\usepackage{comment}
\def\SU{\mathrm{SU}}
\def\MeV{\;\! \mathrm{MeV}}
\def\GeV{\;\! \mathrm{GeV}}

\def\be{\begin{equation}}
\def\ee{\end{equation}}
\def\del{\partial}
\def\eqref{(\ref)}

\begin{document}

\hbox{\small MIT-CTP/4784}

\title[Chiral EFT and the Structure of Hadrons from Lattice QCD]{Chiral Effective Theory Methods and their Application to the Structure of Hadrons from Lattice QCD}

\author[cor1]{P.~E.~Shanahan}
\address{Center for Theoretical Physics, Massachusetts Institute of Technology, Cambridge, MA 02139, U.S.A.}
\eads{\mailto{pshana@mit.edu}}

\begin{abstract}
For many years chiral effective theory (ChEFT) has enabled and supported lattice QCD calculations of hadron observables by allowing systematic effects from unphysical lattice parameters to be controlled. In the modern era of precision lattice simulations approaching the physical point, ChEFT techniques remain valuable tools. In this review we discuss the modern uses of ChEFT applied to lattice studies of hadron structure in the context of recent determinations of important and topical quantities. We consider muon $g-2$, strangeness in the nucleon, the proton radius, nucleon polarizabilities, and sigma terms relevant to the prediction of dark-matter--hadron interaction cross-sections, among others.

\end{abstract}

\section{Introduction}

One of the prime motivators for lattice QCD is its potential to confront experiment in the nonperturbative regime. Its success on this front has historically been tied to chiral effective theory (ChEFT), whose essential role was to bridge the gap between the physical region of light quark masses and simulations with computationally less demanding, heavier, quark masses. In the current era of high-precision lattice studies approaching the physical point, chiral extrapolation techniques remain important. As will be described in the coming sections, the ChEFT formalism has become a refined tool which informs lattice QCD  in both qualitative and quantitative ways and extends the physics impact of state-of-the-art simulations.

In this review we discuss a selection of recent hadron structure results from lattice QCD where the application of ChEFT methods played a key role. In particular, we consider topical issues %
including calculations of muon $g-2$, strangeness in the nucleon, the proton radius, and sigma terms relevant to the prediction of dark-matter--hadron interaction cross-sections. Throughout the discussion we maintain a focus on ChEFT techniques tailored to lattice QCD in the high-precision era. 
Now, as in the past, the power of the lattice-QCD/ChEFT combination comes largely from the facility of both techniques to probe QCD beyond the physical parameter space. In particular, ChEFT provides a framework to extrapolate unphysical lattice simulations to quantities of physical interest. On the other hand, lattice simulations can constrain the universal low-energy constants (LECs) of ChEFT (which, for example, encode the quark-mass dependence of physical quantities). These LECs can then be used to make predictions within the ChEFT formalism of other quantities of interest which were not, and in some cases can not be, directly simulated. %

 After a very brief summary of ChEFT for hadrons in section~\ref{sec:ChEFT}, %
we consider some of the ChEFT approaches used to achieve recent significant physics results.
For the purposes of this review, we divide these approaches into two broad classes: %
\begin{itemize}
\item Extrapolation to the physical point (section~\ref{sec:extrap}): \\
After calculating expectation values of observables at unphysical lattice parameters which are computationally feasible (large quark masses, finite lattice volumes and lattice spacings, discrete values of the three-momentum transfer in the case of form factors, etc.), one extrapolates to the physical point in order to make contact with experiment. This is the traditional and most common use of ChEFT applied to lattice QCD. We consider in particular:
\begin{itemize}
\item Extrapolation in meson masses;
\item Estimation of finite volume (FV) artifacts.
\end{itemize}
\item Access to new quantities through the determination of LECs (section~\ref{sec:LECs}):\\
The goal in this approach is to use lattice QCD simulations to constrain universal LECs of ChEFT and thus make predictions of quantities other than those simulated. As well as the standard leading and next-to-leading order SU(2) and SU(3) LECs, we consider other cases in SU(3) baryon ChEFT where this procedure allows the prediction of quantities closely related to those simulated. These include derivative quantities determined from the slope of lattice results with respect to some parameter (e.g., the proton radius and sigma terms). In addition, one can generally use isospin-averaged ($N_f=2+1$ flavour) lattice simulations to constrain isospin-breaking effects in the simulated quantities, and one can often `unquench' partially-quenched simulation results.
\end{itemize}

\section{A brief introduction to ChEFT}
\label{sec:ChEFT}

The possibility of building a phenomenological effective theory of low-energy QCD exists because there is a mass gap between the pseudoscalar mesons $(\vec{\pi},\vec{K},\eta)$, which are the lightest hadrons, and all other states and resonances. 
This is elegantly explained by the Nambu-Goldstone mechanism: in the limit of vanishing quark mass the pseudoscalar mesons are massless bosons arising from the spontaneous breaking of the chiral symmetry. 
The construction of an effective Lagrangian describing only the low-energy Goldstone-boson modes, but incorporating the full chiral symmetry of QCD, allows a systematic analysis of the implications of the symmetries and symmetry-breaking pattern, with higher-order corrections treatable in the sense of perturbative field theory.

In nature, the octet mesons are only approximately Goldstone because of the {\it explicit} chiral symmetry breaking by the finite quark masses; the quark-mass term in the Lagrangian, $-M_q\overline{\psi}\psi$, is not invariant under chiral transformations. Nevertheless, as the physical QCD vacuum lies very close to a spontaneously broken phase of an exact chiral symmetry, we can treat the explicit breaking as a perturbation about the chiral limit, giving rise to the small masses of the physical octet mesons.

Encoding this expectation, the effective chiral Lagrangian is given by the most general expression of the form
\begin{equation}
\mathcal{L}_{\textrm{{eff.}}} = \mathcal{L}_0 + \mathcal{L}_{\textrm{{SB}}},
\end{equation}
which satisfies the following conditions:
\begin{itemize}
\item{$\mathcal{L}_0$ possesses the same symmetries as the chirally-symmetric part of the QCD Lagrangian. That is, it is invariant under the chiral flavour group $\SU(3)_L \otimes \SU(3)_R$.}
\item{The symmetry group is spontaneously broken to $\SU(3)_V$ by the ground state of the theory.}
\item{The Goldstone modes resulting from the broken symmetry are the only massless, strongly-interacting particles.}
\item{The explicit symmetry-breaking part, $\mathcal{L}_{\textrm{{SB}}}$, is small, can be treated perturbatively, and generates small masses for the pseudo-Goldstone mesons.}
\end{itemize}
By construction this Lagrangian will produce the same low-energy expansion as QCD itself. The systematic framework underpinning that expansion---an ordering in powers of energies and momenta (generically denoted by $p$) of the interacting particles such that any matrix element or scattering amplitude is organized as a Taylor series in $p$---is called chiral perturbation theory or ChEFT.

The ChEFT expansion gives a model-independent description of QCD observables in the low-energy region. Contributions at each successive order are systematically generated by incorporating terms involving higher derivatives and increased powers of the quark masses into the chiral Lagrangian. In addition to the resulting tree-level contributions at each order, loops with interaction vertices taken from the lower-order Lagrangian must be considered, i.e., chiral perturbation theory corresponds to an expansion in both quark-mass and momentum-dependent interactions and increasing loop complexity. Progressively higher-dimension operators are suppressed by higher inverse powers of the chiral-symmetry--breaking scale, $\Lambda_\chi\approx 1\GeV$, which physically corresponds to the range of validity of the effective theory. At any given order, a finite number of {\it a-priori} unknown low-energy constants (LECs) encode the short-distance physics. 

In a practical sense, this formalism provides exactly the framework needed for the extrapolation of lattice simulation results at larger-than-physical quark masses to the physical point. ChEFT expansions of hadronic quantities are closed-form functions of the quark masses, with all dependence explicit, with a finite number of LECs to be determined from the numerical simulations. In fact, this application of ChEFT to lattice QCD has now been in use for 35 years~\cite{Marinari:1981nu,Hamber:1981zn} and has celebrated many successes.

In the baryon, rather than meson, sector, the ChEFT formalism must be somewhat modified. Higher-derivative operators involving baryon fields are not suppressed in the same way as those involving the meson fields (since if $M_B$ denotes the baryon-mass matrix, $M_B/\Lambda_\chi \sim \mathcal{O}(1)$).
This complicates the low-energy structure of the meson-baryon system considerably; there is no longer a one-to-one mapping between the loop and small-momentum expansions. This was the primary technical difficulty with the original, relativistic, formulation of baryon ChEFT in the late 1980s~\cite{Gasser:1987rb}, and it is the main reason that ChEFT in the one baryon sector is significantly less certain than in the meson sector. In response to the difficulties with relativistic baryon ChEFT, heavy-baryon ChEFT was developed, in which baryons are treated as heavy static fermions~\cite{Jenkins:1990jv}. 
This formalism allows a restoration of the chiral order but suffers from the deficiency that it is no longer manifestly Lorentz invariant. This becomes apparent in the analysis of certain form factors where the expected analyticity properties do not emerge~\cite{Becher:1999he}. A second commonly-used variant of baryon ChEFT is the infra-red regularized formalism~\cite{Becher:1999he}, which simultaneously accounts for manifest Lorentz invariance and chiral order.   

In all formulations of baryon ChEFT, the lowest-lying decuplet of spin-$\frac{3}{2}$ baryon resonances plays a particularly important role because of the closeness of the average decuplet mass to the average octet baryon mass; the physical $N$--$\Delta$ mass splitting is $\delta \approx 300\MeV$. In the application of the ChEFT formalism to lattice simulation results, this scale is comparable to relevant values of the pseudo-Goldstone boson mass $m$. As one cannot claim that $m\ll\delta$, it is in general prudent to retain explicit decuplet fields, rather than integrate them out.
Higher baryon resonances are, in general, sufficiently heavy to be consistently integrated out of the low-energy effective theory. Even allowing for unphysically-large meson masses $m\approx 500\MeV$---of a comparable scale to the mass gap between the nucleon and higher $N^*$ resonances---these fields do not necessarily need to be included explicitly but can be mimicked by higher-dimension operators whose effects are of a similar size. For example, the $N(1440)$ lies only $500\MeV$ above the $N(939)$, but it is estimated that the contribution to typical octet baryon amplitudes from this state is no more than $10\%$ that of the $\Delta(1232)$~\cite{Jenkins:1990jv}.
This can be understood physically using an intuitive argument provided by the quark model: the wavefunctions of the octet and decuplet baryons differ only in the arrangement of spin, while higher resonances have different spatial wavefunctions. As the hyperfine spin-spin interaction is relatively weak, it is energetically easier for an octet baryon to be converted into a decuplet baryon than for it to transition to other excited states. 
For these reasons it is now common practice to include the spin-$\frac{3}{2}$ decuplet, but no higher baryon resonances, into the effective chiral theory.

While the modern heavy-baryon and infra-red-regulated approaches to baryon ChEFT have had many successes, the nature of the convergence of these theories, especially in the full SU(3) formalism, is still debated. Famously, the mass of the nucleon shows major deviations from the naive expectations of dimensionally-regulated (DR) ChEFT except in very close vicinity of the chiral limit~\cite{Beane:2014oea,Beane:2004ks,Young:2005tr}. Because of this behavior, there have been a number of efforts to partially resum higher orders in the chiral expansion. In particular, the finite-range regularization (FRR) scheme, which takes into account the extended nature of baryon fields, has had great success. Physically, DR at any fixed order treats meson-baryon couplings as point-like and does not take into account the finite size of the baryon, instead integrating over loop momenta far beyond the scale where the theory has any significance~\cite{Donoghue:1998bs,Lepage2007,Young2005}. 
In general, the incorrect high-energy/short-distance physics included in this way can be absorbed by a redefinition of the LECs appearing in the local Lagrangian. In some cases, however, in particular in SU(3) baryon chiral perturbation theory, the incorrect short-distance physics included in the loops can negatively affect the convergence of the chiral expansion at any finite order. The reason is that the residual (incorrect) short-distance contributions are large even after renormalization (see, e.g., Fig. 4 in Ref~\cite{Donoghue:1998bs}). Large effects can of course still be removed by the adjustment of LECs, but those LECs must consequently also be large. As a result, each term in the expansion is sizeable and it does not clearly converge. If one were able to carry out the process to all orders, one would, of course, still obtain the correct result. However, at any finite order, the incorrect short-distance physics included in the loops has obscured the convergence of the expansion, leaving a formally correct but ineffective procedure.

FRR circumvents this issue by introducing a finite ultraviolet cutoff into loop integrands. This cutoff (i.e., a mass parameter $\Lambda$) physically corresponds to the fact that the source of the meson cloud is an extended structure~\cite{Stuckey1997,ThomasNucl.Phys.Proc.Suppl.119:50-582003,Donoghue1999,Leinweber1999}.
The form of the regulator used, which could for example be chosen to be a sharp cutoff or dipole, does not affect the leading-order non-analytic structure of the expansion~\cite{Leinweber2004}. Furthermore, the renormalization constants may be fixed by matching to lattice simulation results, eliminating dependence on the regulator.
This approach offers improved convergence over dimensionally-regulated SU(3) chiral expansions because the parameter $\Lambda$ remains finite; FRR effectively partially resums the chiral expansion, leaving the long-distance model-independent physics to dominate at the lower orders. In the limit $m_\phi / \Lambda \rightarrow 0$ (where $m_\phi$ denotes the loop meson mass), FRR becomes equivalent to DR. %
It is worthwhile to note here that there is some evidence that, for a given functional form of the regulator, the optimal regularization scale as constrained by lattice results is associated with an intrinsic scale. By examining the renormalization flow of LECs for various nucleon properties such as its mass, magnetic moment and charge radius, Hall and collaborators found a consistent optimal scale at about 1~GeV for a dipole regulator~\cite{Hall:2012pk,Hall:2010ai,Hall:2013oga}. A straightforward interpretation is that this scale characterises the finite size of the nucleon. It would be interesting to see how that analysis extends to include lattice simulations for hyperons both considered individually and fit simultaneously across the baryon octet.

In addition to the use of ChEFT to extrapolate lattice simulation results from larger-than-physical to physical pseudoscalar masses, it has been common since the late 1980s to fit and extrapolate away the finite-volume dependence of lattice data using the same formalism~\cite{HELLER1988189}. This approach takes advantage of the fact that the chiral effective Lagrangian is volume-independent for periodic boundary conditions~\cite{Gasser:1987zq}; the same Lagrangian governs both the quark-mass and volume dependence of observables. 
Intuitively, one understands that on a finite lattice volume the dominant finite-volume effects come from the exchange of mesons `around the world' of the lattice as a result of the periodic boundary conditions; a pion emitted from a nucleon can not only be reabsorbed by the same nucleon, but also by one of the periodic images of the original nucleon which appear at distances of integer multiples of $L$ in each direction. As a consequence, the mass of a hadron, for example, receives corrections of order $e^{-m_\pi L}$ to its asymptotic value. For typical numerical simulations, $m_\pi L\ge 4$ and the finite-volume corrections are small compared to the statistical uncertainties.
In practice, explicit expressions for finite-volume artifacts are written in terms of the loop integrals which represent the meson cloud in the chiral perturbation theory formalism. The finite-volume shift to the value of some observable is modelled by the difference between the loop expression evaluated on a lattice of length $L$---a sum over the discrete allowed momenta which are integer multiples of $2\pi/L$---and the infinite-volume loop integral. This description is applicable as long as the dominant effects arise from the deformation of the pion cloud in the finite volume, i.e., as long as $L$ is not too small.
The accuracy of this model has been confirmed, for the case of the octet baryon masses, by a detailed numerical study using multiple lattice volumes~\cite{Beane:2011pc}.

Lattice discretization effects have also been incorporated into the ChEFT formalism~\cite{Sharpe:1998xm,Rupak:2002sm}. Named lattice chiral perturbation theory (LChPT), the approach allows one to calculate the analytic $a$-dependence of hadron observables simultaneously with the quark-mass dependence; non-polynomial terms in $a$ arise from chiral loops. LChPT thus provides a formalism for chiral and continuum extrapolations. Although fully developed~\cite{Sharpe:1998xm,Lee:1999zxa,Rupak:2002sm,Bar:2004xp}, baryon-sector LChPT is less often used in the analysis of lattice results than continuum ChEFT.
In part, this is because the Symanzik improvement scheme~\cite{Symanzik:1983dc}--the process of improving the lattice action by adding terms which vanish in the continuum limit but act to cancel discretization artefacts at finite-$a$---is routinely used. With any of several methods including $\mathcal{O}(a)$-improved Wilson fermions, staggered fermions, domain-wall fermions, and overlap fermions, the leading cutoff effect is of order $a^2$ which is small relative to the statistical precision for many calculations of baryon observables. Remaining discretization effects can often be absorbed by the addition of simple analytic terms proportional to $a^2$ in the analysis. Another practical consideration is that many lattice studies still include only a single value of the lattice spacing, focussing instead on the control of the (often more significant) finite-volume and chiral extrapolation effects. Of course, this is changing and for the analysis of state-of-the-art calculations which include very precise lattice data, many degrees of freedom, and multiple lattice spacings, LChPT is the natural analysis tool. In the meson sector, where calculations are typically considerably more precise than in the baryon sector, the technique is commonly used.

Finally, we comment that for any lattice studies with close-to-physical parameters, ChEFT provides a valuable check of the systematic uncertainties in the simulation results. By varying the quark mass away from the physical value one might discover that the numerical results do not match the predictions of the continuum effective theory, perhaps because the lattice spacing is still too large, the volume too small, or because of some other systematic effect. In other words, even for modern lattice simulations at the physical pseudoscalar masses, ChEFT provides a useful tool for the validation of results obtained with lattice QCD.
For a more extensive introduction to ChEFT for hadrons aimed at lattice theorists, we refer the reader to Refs.~\cite{Golterman:2009kw,Sharpe:2006pu,Kronfeld:2002pi}.

\section{Extrapolation to the physical point}
\label{sec:extrap}

The extrapolation of lattice QCD observables from unphysical parameter space---most commonly unphysical pseudoscalar masses and finite lattice volume---to the physical point is the prototypical application of ChEFT to the lattice. In this section we consider a selection of recent physics results for a range of hadron structure observables where such extrapolation has either played a key role or where further efforts are of current importance.

\subsection{Nucleon electromagnetic form factors}

The electromagnetic form factors (EMFFs) of the nucleon are hadronic structure observables which encode the fact that the proton and neutron are not point particles, but rather have some extended structure. The Sachs electric and magnetic form factors $G_E$ and $G_M$ describe the spatial distribution of the charge and magnetization density in the nucleon~\cite{Rosenbluth:1950yq} and are expressed as functions of the probing momentum scale, $Q^2$.

The first measurements of proton form factors were reported in 1955~\cite{Hofstadter:1955ae}, followed by the first measurement of the neutron magnetic form factor in 1958~\cite{PhysRev.110.552}. Half a century later, the precise determination of these quantities, and their interpretation within the framework of QCD, remains a defining challenge for hadronic physics research~\cite{Arrington:2006zm}. With ever-improving experimental measurements of the nucleon form factors revealing slight deviations from the phenomenological dipole form~\cite{Bernauer:2010wm,Jones:1999rz,Ron:2011rd,Zhan:2011ji}, it is of renewed importance to calculate precise QCD benchmarks for these functions. In addition to providing such benchmarks, lattice studies also provide an {\it interpretation} of the experimental results for the electromagnetic form factors in the context of QCD. For example, the simulations give general insight into the environmental sensitivity of the distribution of quarks inside a hadron~\cite{Leinweber:1990dv,Boinepalli:2006xd} by discriminating between different quark-flavour contributions to the form factors. The lattice method can also reveal the dependence of these quantities on quark mass~\cite{Cloet:2002eg,Thomas:2002sj,Shanahan:2012wh} and allows a separation of quark-line--connected and disconnected terms~\cite{Sharpe:2001fh,Savage:2001dy,Leinweber:2002qb,Tiburzi:2009yd,Allton:2005fb}, providing both a great deal of physical insight and valuable information for model-building~\cite{Cloet:2013gva}.

With the majority of lattice simulations of the EMFFs performed at larger-than-physical values of the pseudoscalar masses, ChEFT techniques play an important role in the extrapolation of simulation results to the physical point. This is especially true in the case of the Sachs electric form factor $G_E$, for which ChEFT predicts rapid change towards the chiral regime. That is, the result of simulations very close to the physical pseudoscalar masses can, and are expected to, differ significantly from those at the physical point. A precise determination of the electric Sachs form factors of the nucleon from the lattice is of particular interest at this time because of the unresolved `proton radius puzzle': the 7$\sigma$ difference between the charge radius of the proton (related to the slope of the form factor $G_E$ in the static limit) as determined from electron-proton scattering experiments~\cite{2012RvMP841527M,Bernauer:2010wm} and from atomic spectroscopy of muonic hydrogen~\cite{Antognini:1900ns}.

\begin{figure}
\centering
\begin{subfigure}[This figure from Ref.~\cite{Alexandrou:2015yqa} shows results from simulations with $N_f =2+1+1$ (filled blue squares)~\cite{Alexandrou:2013joa} and $N_f =2$~\cite{Alexandrou:2011nr}(filled red circles) for a pion mass of about $300\MeV$. Also shown are results with $N_f =2+1$ domain wall fermions at $m_\pi = 297\MeV$ (crosses)~\cite{Syritsyn:2009mx}, with a hybrid action with $N_f =2+1$ staggered sea and DWF at $m_\pi = 293\MeV$ (open orange circles)~\cite{Bratt:2010jn}, and with $N_f =2$ clover at $m_\pi = 290\MeV$ (asterisks)~\cite{Capitani:2010sg}. The dashed line shows the parametrization of the experimental data from a number of experiments as given in Ref.~\cite{Kelly:2004hm}.]{\label{fig:GE1}
\includegraphics[width=0.6\textwidth]{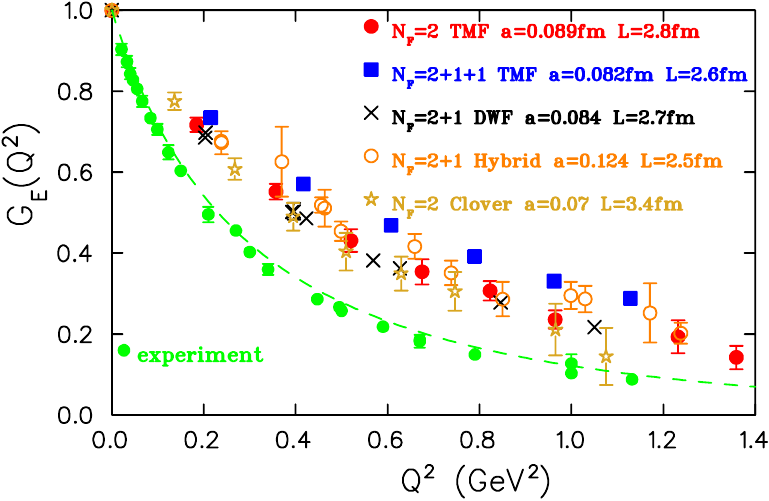}
}
\end{subfigure}
\begin{subfigure}[This figure is updated from Ref.~\cite{Shanahan:2014cga}. The points show the results of a chiral extrapolation of sets of lattice simulations with the lightest pion masses around $310\MeV$ (blue circles) and $265\MeV$ (green crosses). The red line is a parameterization of experimental results from Ref.~\cite{Kelly:2004hm}.]{\label{fig:GE2}
\includegraphics[width=0.45\textwidth]{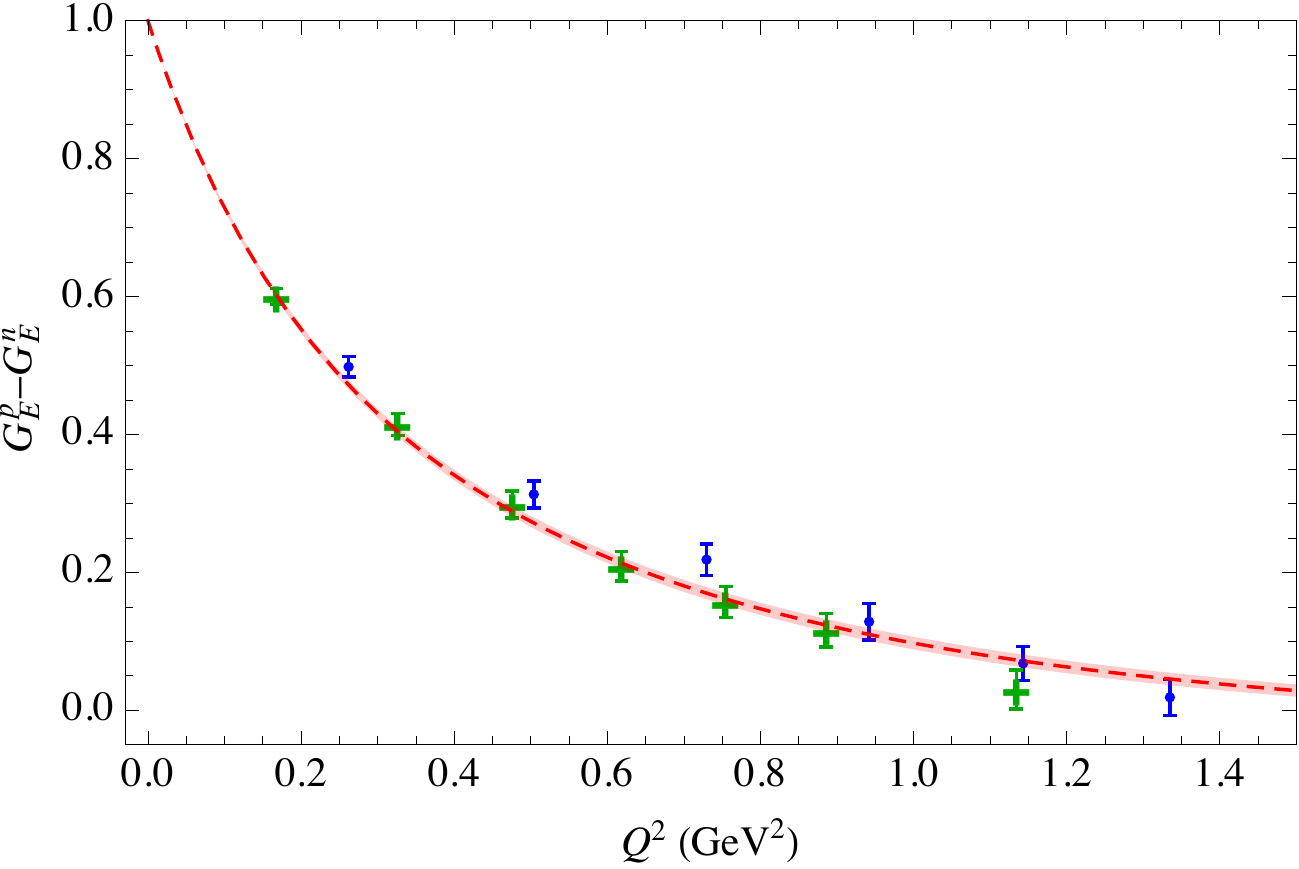}
}
\end{subfigure}
\begin{subfigure}[This figure from Ref.~\cite{Green:2014xba} shows near-physical simulations with a pion mass of around $149\MeV$. The solid line depicts a parameterization of experimental results.]{ \label{fig:GE3}
\includegraphics[width=0.49\textwidth]{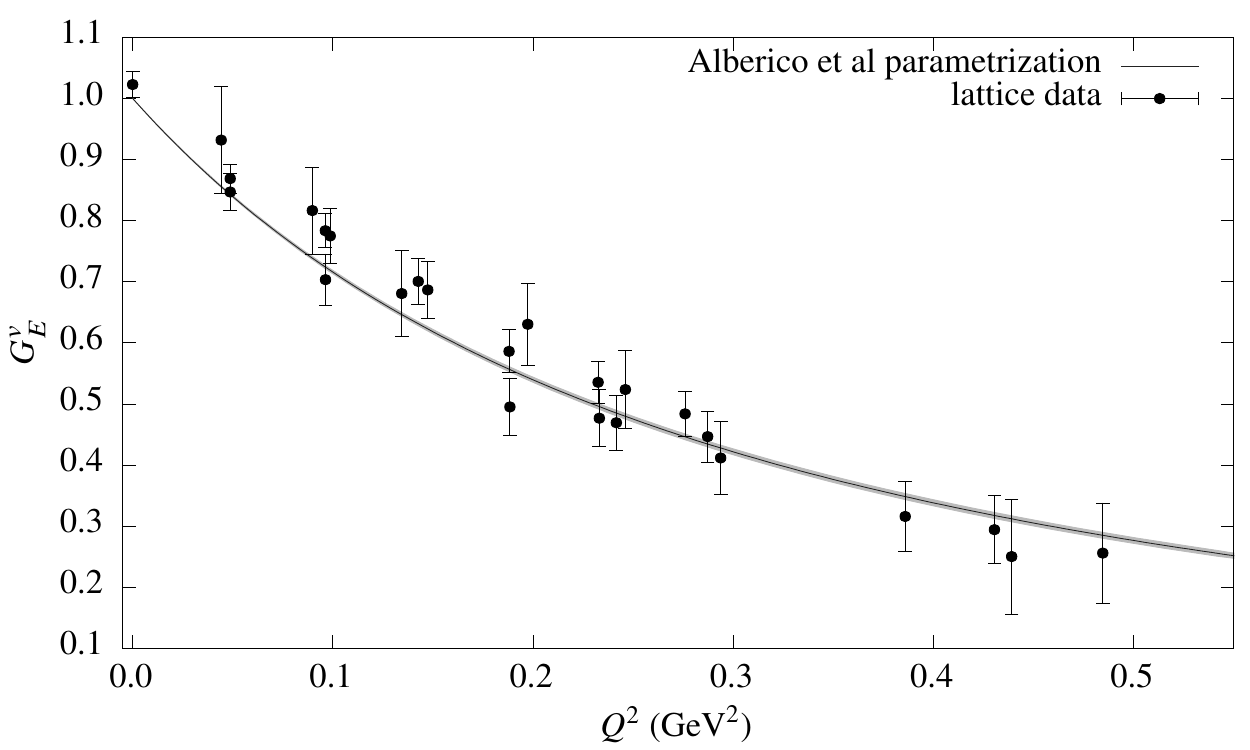}
}
\end{subfigure}
\caption{\label{fig:GE}The $Q^2$-dependence of the isovector combination of electric Sachs form factors, $G_E=(G_E^p-G_E^n)=G_E^v$, from a number of recent lattice QCD simulations.}
\end{figure}

Over the last several years many lattice collaborations have presented increasingly precise simulation results for the EMFFs~\cite{Green:2014xba,Capitani:2015sba,Bratt:2010jn,Syritsyn:2009mx,Alexandrou:2013joa,Alexandrou:2007xj,Alexandrou:2006ru,Gockeler:2003ay,Hagler:2007xi,Lin:2008uz,Liu:1994dr,Sasaki:2007gw,Hagler:2009ni,Boinepalli:2006xd,Yamazaki:2009zq,Collins:2011mk,Lin:2008mr,Wang:2008vb,Shanahan:2014uka,Shanahan:2014cga}. Here we highlight just a few recent results for $G_E$. Figure \ref{fig:GE1}, taken from Ref.~\cite{Alexandrou:2015yqa}, displays results for the isovector electric Sachs form factor, i.e., $G_E^p-G_E^n$ where $p$ and $n$ denote the proton and neutron form factors, against the probing momentum scale $Q^2$. This combination is of particular interest since most lattice simulations currently omit disconnected quark line contributions (although recent progress towards the calculation of disconnected terms is made in, for example, Ref.~\cite{Green:2015wqa}). The omitted terms should cancel in the isovector combination given charge symmetry, yielding a quantity that can be directly compared with experiment. At larger-than-physical values of the pion mass around $300\MeV$, calculations with $N_f =2+1$ domain wall fermions~\cite{Syritsyn:2009mx}, $N_f =2$ Wilson improved clover fermions~\cite{Capitani:2010sg} and a hybrid action~\cite{Bratt:2010jn} are in good agreement~\cite{Alexandrou:2013joa} but deviate systematically from the experimental values. This behavior can be understood quantitatively using ChEFT extrapolation techniques. Figure~\ref{fig:GE2} taken from Ref.~\cite{Shanahan:2015caa}, shows the results of a range of $N_f=2+1$ flavor lattice simulations, with the lightest pion masses around $310\MeV$ and $265\MeV$ for the two sets of simulations shown (blue circles and green crosses respectively), extrapolated to the physical point using a finite-range regulated ChEFT formalism which includes decuplet baryon resonances. Clearly, there is excellent agreement with experiment. Extrapolated instead to a pion mass around $300\MeV$, these simulations agree well with those shown in Fig.~\ref{fig:GE1}.  Similar results were obtained in Ref.~\cite{Capitani:2015sba}, where chiral extrapolation of lattice simulation results, including attention to excited state contamination (which becomes more important as the pion mass is reduced), yielded good agreement with experiment. The authors of both Refs.~\cite{Capitani:2015sba} and \cite{Shanahan:2015caa} comment that it was important to this agreement to extrapolate the form factors themselves~\cite{Wang:2007iw}, and avoid the systematic uncertainties inherent in the use of a dipole fit in $Q^2$ before extrapolation. 

Recently, lattice simulations very close to the physical point, with $146\MeV$ pions, have been presented in Ref.~\cite{Green:2014xba}. Shown in Fig.~\ref{fig:GE3}, these direct simulations are also consistent with experiment, and with the extrapolated results shown in Fig.~\ref{fig:GE2}. Clearly, the disagreement between lattice simulations with pion masses of order $300\MeV$ and experimental values for $G_E$ can be understood in the context of chiral perturbation theory, and this understanding is supported by near-physical-point simulations.

Given the consistency between experimental values and lattice simulation results for $G_E$, there is hope that with improved precision---most importantly, lower values of $Q^2$---such simulations will eventually be able to provide a precise value of the charge radius from QCD. One way to overcome the restriction to lattice quantized momenta and reach the smaller values of $Q^2$ needed for a precise extraction of the charge radius is to impose twisted boundary conditions on the quark fields~\cite{Bedaque:2004kc,deDivitiis:2004kq,Sachrajda:2004mi}.
Since computational restrictions currently limit simulations to partially-twisted boundary conditions, however, lattice results obtained at different values of twist angles are correlated and this method does not immediately reduce the statistical uncertainty on the charge radius compared to the more traditional approach. Recently a new method has been proposed which allows the charge radius itself to be computed directly at zero momentum~\cite{deDivitiis:2012vs}. With a ChEFT analysis of the finite-volume effects inherent in this new method presented in Ref.~\cite{Tiburzi:2014yra}, this approach seems to be a promising way forward.

\subsection{Nucleon polarizabilities}

The polarizabilities of the nucleon parametrize the deformation of its charge and magnetization distributions in external electric and magnetic fields. In other words, these observables describe how easily electromagnetic interactions induce transitions to low-lying excited states. They encode information about the symmetries of the nucleon as well as the strength of interaction of its constituents with each other and with the photon. As well as electric and magnetic polarizabilities $\alpha_{E1}$ and $\beta_{M1}$, a spin-half object like the nucleon has four spin-polarizabilities, denoted $\gamma_i$, $i=1\ldots 4$, which encode the object's spin-dependent response to an electromagnetic field.
The polarizabilities are of particular interest at this time; they play an important role in the Lamb shift of muonic Hydrogen, which is the least-known ingredient of the proton-radius puzzle, as well as in radiative corrections to the proton charge radius, and constitute the biggest source of uncertainty in theoretical determinations of the proton-neutron mass shift. 

Over the last several years a number of new results have been published from experiments devoted to understanding the nucleon polarizabilities. Results from both MAXlab~\cite{Myers:2014ace,Myers:2015aba} and MAMI~\cite{Martel:2014pba} were published within the last year. In parallel, there have been considerable efforts to determine the nucleon polarizabilities theoretically from QCD, including a number of lattice QCD simulations~\cite{Chang:2015qxa,Lujan:2014qga,Primer:2013pva,Hall:2013dva,Engelhardt:2007ub,Freeman:2014kka,Detmold:2010ts,Engelhardt:2010tm}. Since all existing simulations have pion masses significantly larger than the physical value, chiral extrapolation formalisms are of pressing interest in particular because the polarizabilities are very sensitive to infrared physics and their mass and volume dependence is considerably stronger than that expected for hadron masses and magnetic moments~\cite{Detmold:2006vu}. While the exploration of nucleon polarizabilities was a natural early application of ChEFT in the baryonic sector and dates back to the early 1990s~\cite{Bernard:1991rq,Bernard:1995dp}, there has been recent progress and work particularly targeted at the extrapolation of lattice QCD simulation results. 

\begin{figure}
\centering
\begin{subfigure}[Figure from Ref.~\cite{Griesshammer:2015ahu}. Comparison between ChEFT predictions and lattice QCD computations for $\alpha_{E1}$. Lattice results from Ref.~\cite{Lujan:2014qga} (blue upward triangles (neutron)), Ref.~\cite{Detmold:2010ts} (red cross (proton) and blue plus (neutron)), and Refs.~\cite{Engelhardt:2007ub,Engelhardt:2010tm} (blue downward triangles (neutron)).]{\label{fig:alphan}
\includegraphics[width=0.48\textwidth]{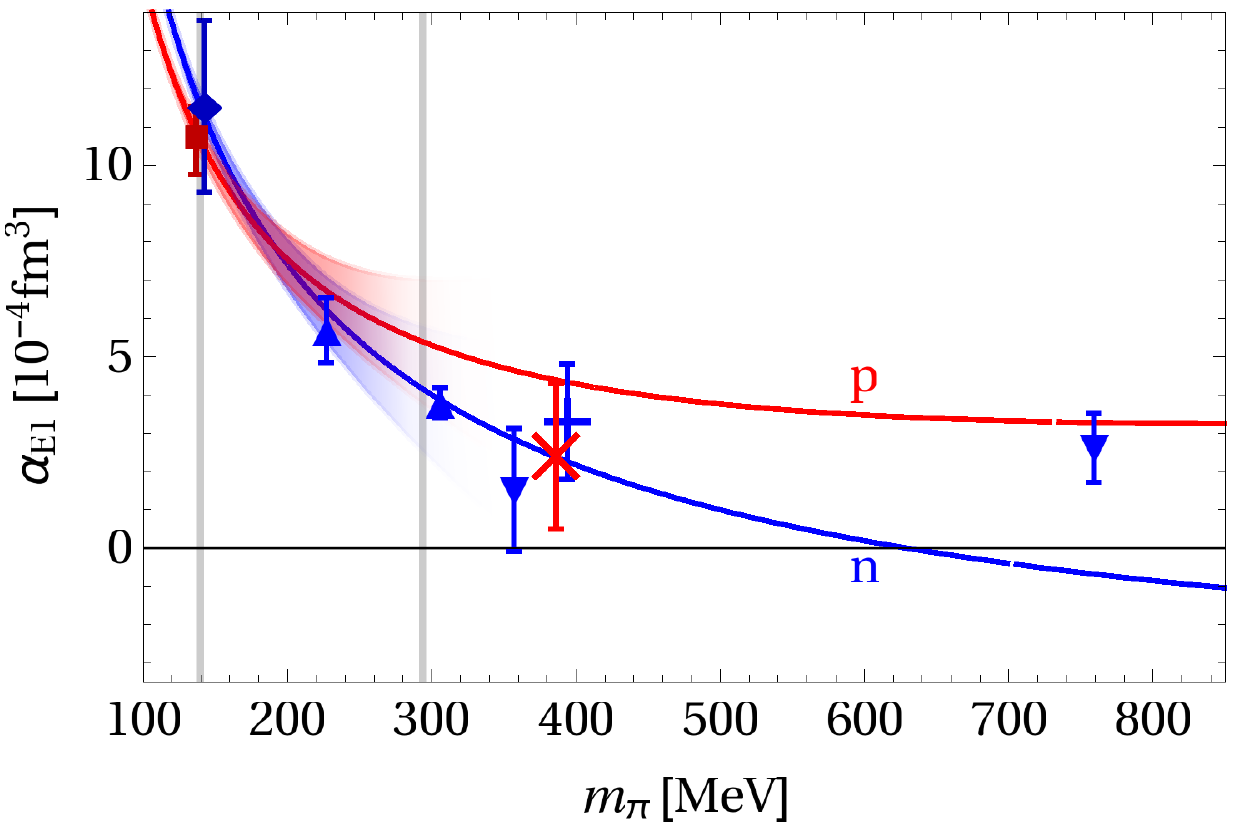}
}
\end{subfigure}
\begin{subfigure}[Figure from Ref.~\cite{Hall:2013dva}. Chiral extrapolation of lattice simulation results (pink cross) for the neutron magnetic polarizability $\beta_{M1}$. compared with experimental results from Refs.~\cite{pdg,Griesshammer:2012we,Kossert:2002jc,Kossert:2002ws}.]{\label{fig:began}
\includegraphics[width=0.46\textwidth]{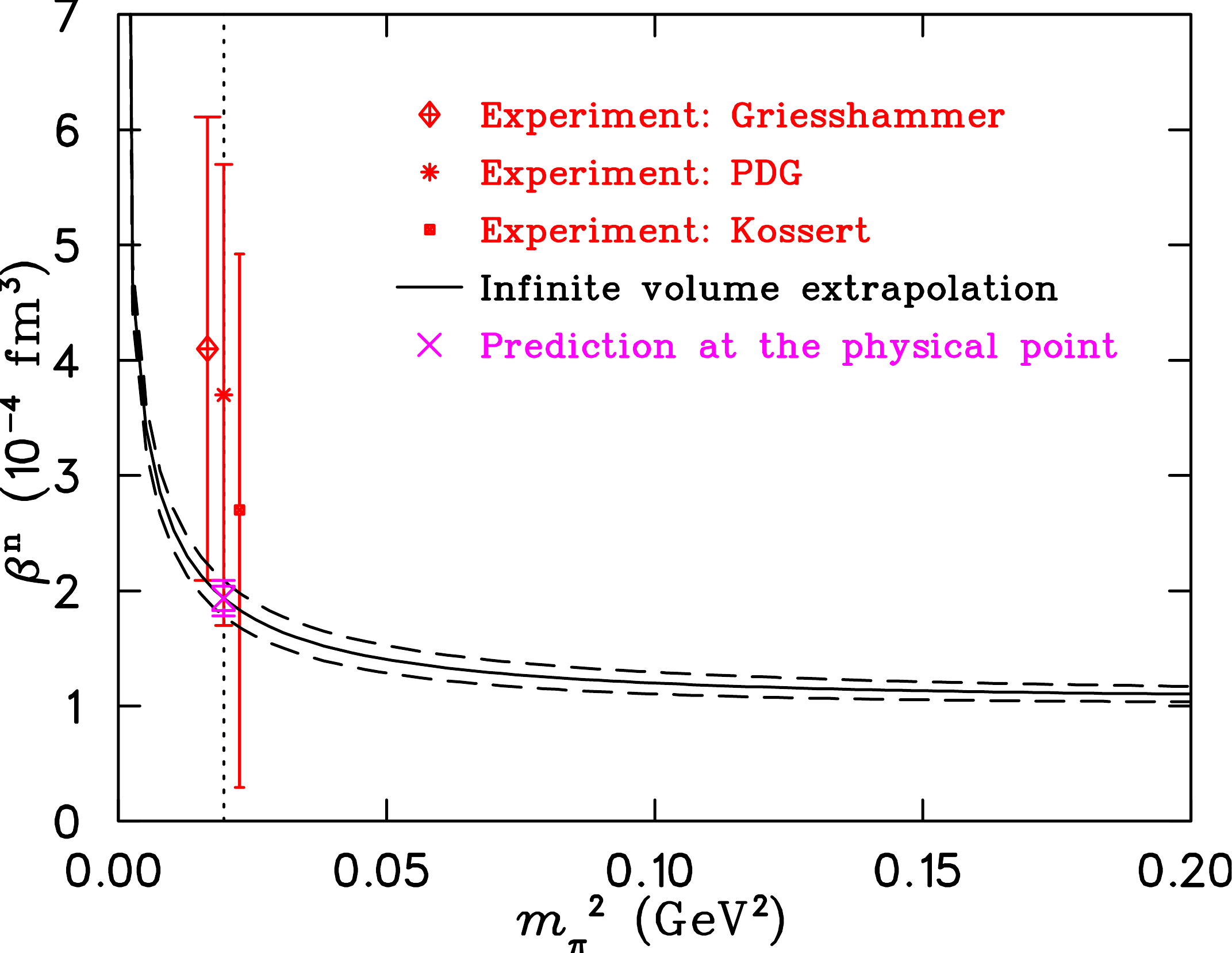}
}
\end{subfigure}
\caption{\label{fig:polfig} Comparison between ChEFT predictions and lattice simulation results for the electric polarizabilities of the proton and neutron (left panel), and comparison between chirally-extrapolated lattice simulation results and experimental values for the neutron magnetic polarizability (right panel).}
\end{figure}

Figure~\ref{fig:polfig} shows some recent lattice and ChEFT results for the nucleon polarizabilities. Shown in Fig.~\ref{fig:alphan}, the dimensionally-regulated chiral perturbation theory formalism (with decuplet degrees of freedom) and careful error analysis presented in Ref.~\cite{Griesshammer:2015ahu} agrees very well with emerging lattice computations, even beyond the mass range of $m_\pi\lesssim 300\MeV$ over which the authors argue that their ChEFT is applicable. With more lattice simulations at light pion masses within the range of applicability of the theory, this formalism will allow a controlled chiral extrapolation of the polarizabilities. A new analysis of finite-volume effects in such lattice simulations, using the framework of heavy-baryon ChEFT, was presented in Ref.~\cite{Hall:2013dva}, where it was noted that box sizes of approximately 7 fm are required to achieve results within 5\% of the infinite-volume results at the physical pion mass. Clearly, future lattice simulations face a trade-off between lighter masses and larger volumes in order to make physical predictions for the nucleon polarizabilities. A first chiral extrapolation of lattice results for the nucleon magnetic polarizability, where ChEFT methods were also used to estimate finite-volume effects and to correct for omitted sea-quark loop contributions, is shown in Fig.~\ref{fig:began}, taken from Ref.~\cite{Hall:2013dva}. These promising results bode well for the future of lattice simulations of this quantity.

\subsection{Hyperon vector form factors}

The Cabibbo-Kobayashi-Maskawa (CKM) matrix elements are fundamental Standard Model parameters which encode the flavor structure of the quark sector. A stringent test of CKM unitarity~\cite{Cirigliano:2009wk,Antonelli:2009ws} is given by the first-row relation $|V_{ud}|^2+|V_{us}|^2+|V_{ub}|^2=1$, where $|V_{us}|$ contributes the largest uncertainties. Determinations of $|V_{us}|$ have traditionally been based on kaon semileptonic and leptonic decays and the hadronic decays of tau leptons. These extractions are in slight tension~\cite{pdg}, although a resolution has recently been proposed~\cite{Boyle:2013xya,MaltmanPrep2,Maltmanprep}.
For the last decade~\cite{Guadagnoli:2004qw} there has been considerable interest in a determination of $|V_{us}|$ from semileptonic hyperon decays studied on the lattice. The hope is that this approach will lead to an improved determination of the $u$-$s$ CKM matrix element independent of extractions from kaon and tau decays. 
Since the product $|V_{us}f_1(Q^2=0)|$ can be extracted from experiment at the percent-level~\cite{Cabibbo:2003cu}, the required lattice input is a precise calculation of the hadronic corrections to the vector form factors $f_1(Q^2=0)$.
In particular, while the Ademollo-Gatto theorem~\cite{PhysRevLett.13.264} protects the vector form factors from leading SU(3)-symmetry--breaking corrections generated by the mass difference of the strange and nonstrange quarks, a quantitative understanding of the second-order corrections to $f_1(Q^2=0)$ is crucial to obtain a precise value of $|V_{us}|$~\cite{Cabibbo:2003cu,Mateu:2005wi}. 

A puzzle that has endured over the last decade is that the sign of the SU(3) breaking corrections determined in quenched and unquenched lattice QCD~\cite{Guadagnoli200763,PhysRevD.79.074508,PhysRevD.86.114502} (at this stage away from the physical pseudoscalar masses) and quark models~\cite{Donoghue:1986th,Schlumpf:1994fb} is, in general, opposite to that determined from relativistic and heavy baryon chiral perturbation theory~\cite{Villadoro:2006nj,Lacour:2007wm,Geng:2014efa,Anderson:1993as,Kaiser:2001yc}  and $1/N_c$ expansions~\cite{PhysRevD.58.094028}. 
The crucial issue faced by lattice determinations of $f_1(Q^2=0)$ is then the accuracy in the extrapolation to the physical point, for which a sound understanding of the ChEFT expansion is essential.

Recent work has shown that finite-volume effects are relatively small for typical lattice simulation parameters with $m_\pi L \ge 4$~\cite{Geng:2014efa}, but that chiral extrapolation needs to be performed more carefully. It has also been emphasized that the order of the chiral extrapolation, finite-volume corrections, and extrapolation in $Q^2$---from $Q^2=-(M_{B_1}-M_{B_2})^2$ which is accessible to simulations with fixed sink momentum to $Q^2=0$---is important. 
Moreover, performing that small shift in $Q^2$ is in general highly dependent on the approach used~\cite{Shanahan:2015dka}.

The global picture from the most recent lattice studies is that the sign of the SU(3)-symmetry-breaking corrections found is consistent with the results of quark models but opposite to that of ChEFT approaches. After attempts at chiral extrapolation, the size of the breaking is generally larger than in other approaches. That is, the discrepancy between lattice and ChEFT predictions remains. Since ChEFT is important to the interpretation of lattice simulations, further theory studies are needed to fully understand it before a reliable lattice-informed extraction of $|V_{us}|$ can be performed based on hyperon semileptonic decays. 
On the lattice side, simulations at a range of light quark masses will of course ameliorate the reliance on ChEFT extrapolations.
Perhaps computationally easier and similarly important, however, are calculations exploring the $Q^2$ range between $Q^2=0$ and the typical value of $Q^2=-(M_{B_1}-M_{B_2})^2$, possibly achieved using boosted systems or twisted boundary conditions.

 \subsection{Hadronic vacuum polarization}

The anomalous magnetic moment of the muon, defined as $a_\mu = (g-2)/2$, is one of the most precisely measured physical quantities. As such, the comparison between experimental and theory values is important in the search for indirect evidence of new physics beyond the mass range directly accessible at the Large Hadron Collider. For a number of years, however, there has been a persistent three to four sigma discrepancy between these values~\cite{Blum:2013xva,Miller:2012opa}. This has motivated extensive experimental and theory efforts aimed at understanding the discrepancy. 
The theoretical error is dominated by hadronic contributions since, in contrast to electroweak quantities, QCD observables cannot be reliably calculated using perturbation theory. Since the lowest-order hadronic contribution is estimated using a dispersion relation which relies on experimental data, a lattice QCD determination of this quantity is extremely desirable and a number of groups have risen to the challenge of such a calculation~\cite{Aubin:2006xv,Boyle:2011hu,DellaMorte:2011aa,Burger:2013jya,Aubin:2015rzx,DellaMorte:2016izp,Chakraborty:2016mwy}. 

ChEFT plays a significant role in the extraction of physically-relevant results from lattice simulations of the hadronic vacuum polarization term. In addition to a careful treatment of finite-volume effects and the light quark mass extrapolation, a precise fit to the low-$Q^2$ region is essential to extract a precise value of $(g-2)$; the leading-order hadronic contribution to this quantity can be expressed as an integral over Euclidean $Q^2$ of the vacuum polarization function. Typically, polynomial fits, continuous forms motivated by models of vector dominance~\cite{Boyle:2011hu,Burger:2013jya} and fits based on staggered chiral perturbation theory coupled to photons~\cite{Aubin:2006xv} are used to parameterize the $Q^2$-dependence of the simulation results (which are, of course, at discrete values of $Q^2$ on the finite simulation volumes) and perform the required integral. While the latter approach is perhaps most rigorously motivated, it is found that it does not represent lattice simulation results; to fit the data well requires the inclusion of the vector particles through resonance chiral perturbation theory~\cite{Aubin:2006xv}. At fixed lattice spacing the non-locality of rooted staggered fermions may also be a cause for concern with this method. %

It has been recently pointed out that a simple trapezoid-rule numerical integration of current lattice data is good enough to produce a result with a less-than-1\% error for the contribution to $(g-2)$ from the interval above $Q^2\ge 0.1$--$0.2$~GeV$^2$~\cite{Golterman:2014wfa}. It is then the low-$Q^2$ region, with $Q^2\le 0.2$~GeV$^2$ that requires the most attention in order to reach the desired goal of sub-percent precision in the hadronic vacuum polarization contribution to the muon $(g-2)$. 
Improvement to ChEFT approaches to this low-momentum regime---noting that the extrapolation can be limited to the low-$Q^2$ region alone with the higher-$Q^2$ region treated by numerical integration---call for the inclusion of $\mathcal{O}(p^4)$ terms.

\begin{figure}
\centering
\includegraphics[width=0.73\textwidth]{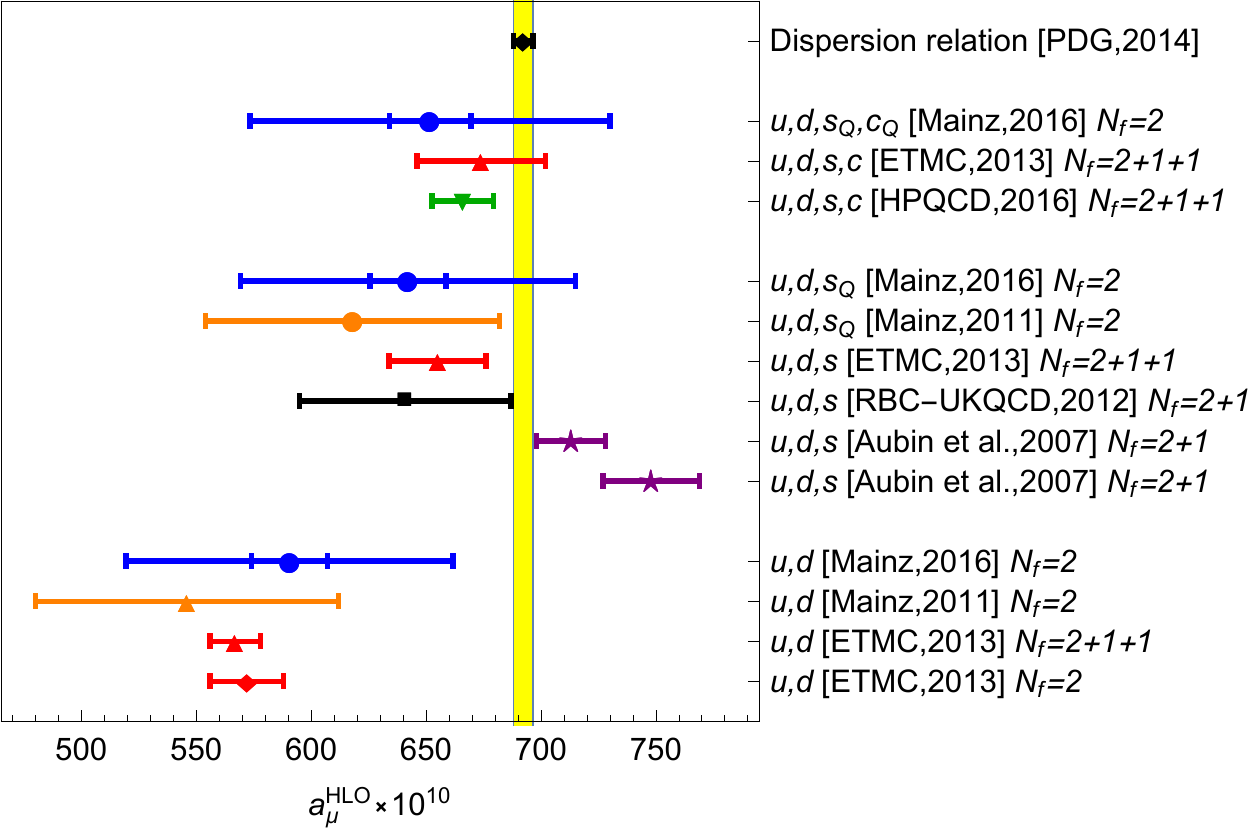}
\caption{\label{fig:gm2}Figure adapted from Ref.~\cite{DellaMorte:2016izp} [Mainz,2016] showing a summary of lattice QCD results for the leading order hadronic contribution to $a_\mu$. Other data points are from Refs.~\cite{Burger:2013jya} [ETMC, 2013], \cite{DellaMorte:2011aa} [Mainz, 2011], \cite{Aubin:2006xv} [Aubin et al., 2007]. The vertical band highlights the dispersion relation result~\cite{pdg} [PDG, 2014].}
\end{figure}

Recently a systematic study of the finite-volume effects in lattice simulations of the hadronic vacuum polarization was made~\cite{Aubin:2015rzx}. Encouragingly, even though leading-order chiral perturbation theory does not provide a good description of the hadronic vacuum polarization, it gives a reasonably good representation of finite-volume effects. These effects cannot be ignored when the aim is a few percent level accuracy for the leading-order hadronic contribution to the muon anomalous magnetic moment, even when using ensembles with $m_\pi L \ge 4$ and pion masses approaching the physical point.

As well as the $Q^2$-extrapolation of lattice data, the chiral extrapolation is also difficult, in particular because the two pion threshold may cause non-linearities and large volume effects. This can be addressed by simulating below the threshold but it is still important to consider including higher order terms in the chiral expansion. A recent simulation with physical-mass light quarks has circumvented this issue~\cite{Chakraborty:2016mwy}. With a careful analysis including multiple values of the lattice spacing and multiple lattice volumes, this calculation supports the $3\sigma$ discrepancy between the experimental and lattice determinations of $a_\mu$~\cite{Chakraborty:2016mwy}. One of the largest uncertainties in that calculation, other than effects that will be mitigated with smaller lattice spacings, arises from the mass-degenerate light quarks used in the simulations. Clearly, new simulations with $m_u\ne m_d$ are a next goal. If the lattice simulations were well-described by a ChEFT form however, perhaps the isospin-breaking uncertainty could be reduced by an approach similar to that described in Section~\ref{sec:CSV} of this review. A summary of the most recent lattice simulation results for $a_\mu$ is given in Fig.~\ref{fig:gm2}.

\section{New quantities through determination of LECs}
\label{sec:LECs}

In addition to the extrapolation of lattice simulations performed in an unphysical region of parameter space to the physical point, ChEFT relates different observables through the symmetries of QCD which are naturally encoded in the ChEFT formalism. In that way, simulations of one set of observables on the lattice can give information about related, but different sets of observables as well. In this section we discuss several important recent examples of the use of ChEFT in this manner. In particular, we describe how the meson-mass--dependence of the octet baryon masses gives information on the nucleon strangeness content as relevant to the interpretation of dark matter direct detection experiments, how isospin-averaged quantities can given information on isospin-breaking effects, and how the ChEFT formalism can describe the relationship between partially-quenched and unquenched, and between  connected and full, lattice simulations.

\subsection{Nucleon sigma terms and strangeness content}

The sigma terms of a baryon $B$ are defined as scalar form factors, evaluated in the limit of vanishing momentum transfer. These quantities provide a measure of quark contributions to the baryon masses and are a key theoretical ingredient for the interpretation of dark matter direct-detection experiments~\cite{Ellis2008,Giedt2009,Ellis2012}. For each quark flavor $q$ and baryon $B$, they are defined by
\begin{align}\label{eq:FH}
\sigma_{Bq} & \equiv m_q \langle B | \overline{q} q | B \rangle = m_q \frac{\del M_B}{\del m_q},
\end{align}
where the last equality is the statement of the Feynman-Hellmann relation in this context. The Feynman-Hellmann theorem relates the derivative of the energy of a system, with respect to some parameter, to the expectation value of the derivative of the Hamiltonian with respect to the same parameter. Here this relation is used to express the sigma terms as derivatives of baryon mass with respect to quark mass~\cite{Feynman1939}.
Clearly, given closed-form ChEFT expressions for baryon mass $M_B$ as a function of the meson masses (related to quark masses by the Gell-Mann-Oakes-Renner relation), fit to lattice simulation results for the baryon masses, the scalar form factors can be evaluated by simple differentiation. 

This method has a considerable advantage over the direct calculation of the sigma terms in lattice QCD; it does not require the evaluation (or estimation) of contributions from quark-line--disconnected diagrams which are represented by noisy and expensive `all-to-all' propagators on the lattice. However, it also has a disadvantage; the application of the Feynman-Hellmann relation requires taking a partial derivative with respect to quark mass. That is, all other parameters must be held fixed, including the strong coupling $\alpha$ (or, equivalently, $\Lambda_{\rm QCD}$). In lattice QCD, there is an apparent ambiguity as to how to define a fixed renormalized coupling $\alpha$~\cite{Shanahan:2013cd,De:2008xt}. This is precisely the issue of lattice scale setting---while lattice simulation results extrapolated to the physical point must be independent of scale-setting scheme, derivative quantities, by definition, make reference to the scale away from the physical point and hence their values may depend on the scheme chosen. Extractions of the strange sigma term in particular seem vulnerable to such effects~\cite{Shanahan:2012wh,Durr:2015dna}. Furthermore, typical lattice trajectories in light-strange quark mass space, with the strange quark mass held essentially fixed as the light quark mass is varied, often do not allow a large enough lever arm for a precise extraction of the strange sigma term~\cite{Durr:2015dna}. This can be overcome by considering different trajectories in the light-strange quark mass plane~\cite{Shanahan:2012wh}.

\begin{figure}
\centering
\begin{subfigure}[$\sigma_{\pi N} = \frac{m_u+m_d}{2}\langle N| \overline{u}u + \overline{d} d | N \rangle$]{
\includegraphics[width=0.47\textwidth]{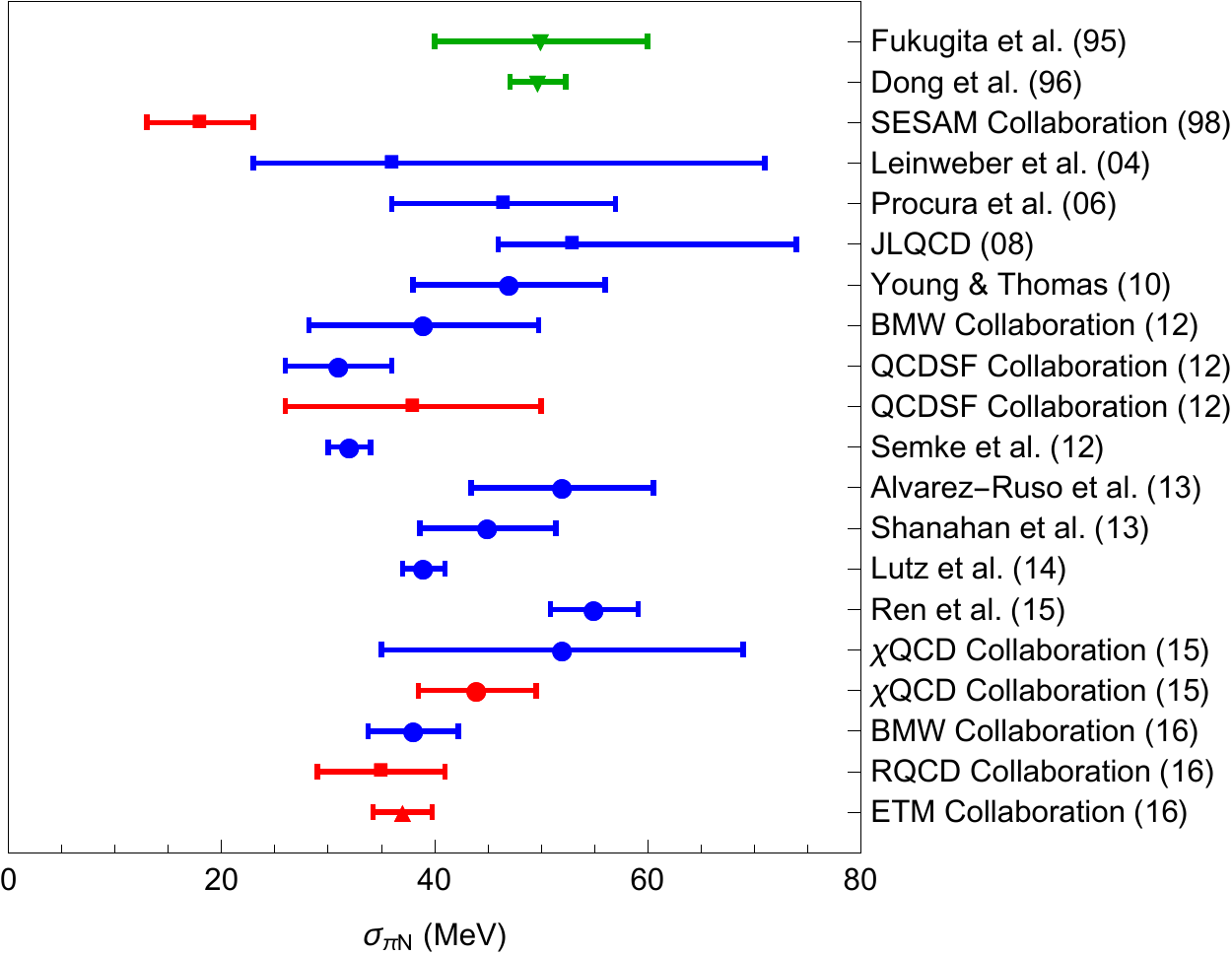}
}
\end{subfigure}
\begin{subfigure}[$f_s^N = \frac{\sigma_{sN}}{m_N}$]{
\includegraphics[width=0.47\textwidth]{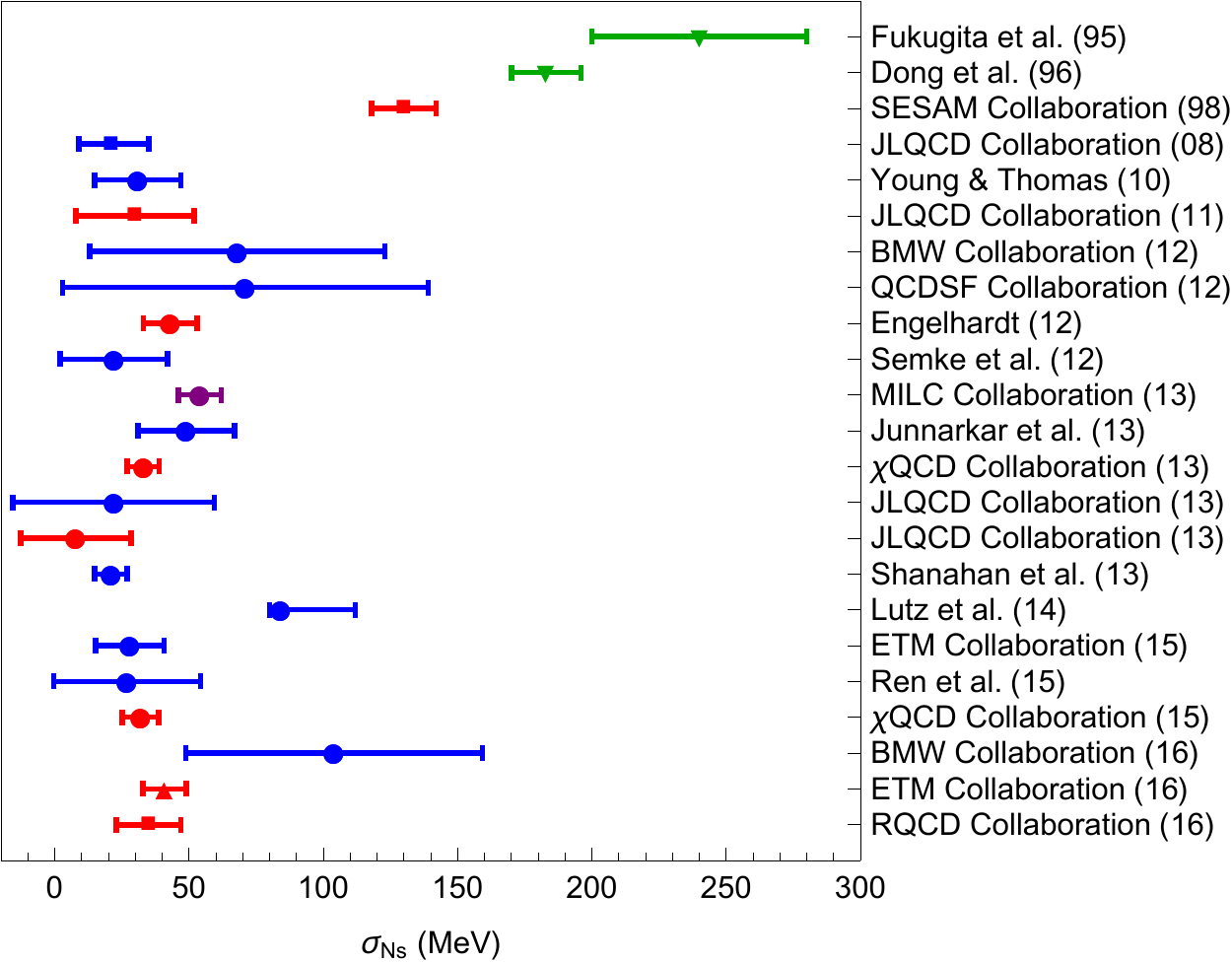}
}
\end{subfigure}
\caption{\label{fig:sigmas}Comparison of some of the lattice results for the sigma terms obtained over the last two decades. Only results which include some attempt at chiral extrapolation (using any formalism), or were simulated at the physical point directly, are shown. Red, blue and purple colours denote direct, Feynman-Hellmann and hybrid approaches, while the green points are from early $N_f=0$ calculations. Squares, circles and upward triangles denote $N_f=2$, $2+1$ and $2+1+1$ studies. Results are from Refs.~\cite{Bali:2016lvx} (RQCD), \cite{Abdel-Rehim:2016won} (ETM), \cite{Durr:2015dna,Durr:2011mp} (BMW), \cite{Yang:2015uis,Gong:2013vja} ($\chi$QCD), \cite{Ren:2014vea} (Ren et al.), \cite{Alexandrou:2013nda} (ETM), \cite{Lutz:2014oxa} (Lutz et al.), \cite{Shanahan:2012wh} (Shanahan et al.), \cite{Oksuzian:2012rzb,Takeda:2010cw,Ohki2008} (JLQCD), \cite{Junnarkar:2013ac} (Junnarkar et al.), \cite{Freeman:2012ry} (MILC), \cite{Semke:2012gs} (Semke et al.), \cite{Engelhardt:2012gd} (Engelhardt), \cite{Horsley:2011wr,Bali:2011ks} (QCDSF), \cite{Young2010} (Young \& Thomas), \cite{Gusken:1998wy} (SESAM), \cite{Dong:1995ec} (Dong et al.), \cite{Fukugita:1994ba} (Fukugita et al.), \cite{Alvarez-Ruso:2013fza} (Alvarez-Ruso et al.), \cite{Procura2006} (Procura et al), \cite{Leinweber:2003dg} (Leinweber et al.).  }
\end{figure}

Recently, the first direct calculation of the sigma terms with dynamical fermions and a physical value of the pion mass was presented~\cite{Yang:2015uis}. The simulations performed in that work allowed physical results to be extracted by an interpolation in the meson masses, rather than a chiral extrapolation, for the first time. Excellent agreement was found between the direct results and those obtained using the Feynman-Hellmann method applied to the same data. These results, which are the most precise to date, are shown alongside the results of previous studies in Figure~\ref{fig:sigmas}. %

In general terms, the results of modern lattice calculations of the sigma terms are in excellent agreement, despite the different approaches used to generate them. These approaches include applications of the Feynman-Hellmann theorem (with various scale-setting schemes) and direct methods, with a range of chiral and volume extrapolation formalisms used to control systematic effects. Of course, different lattice parameters and actions are also used.  Shown in Figure~\ref{fig:sigmas}, the calculations indicate a value for the light-quark sigma term of $\sigma_{\pi N} = m_l \langle N | \overline{u} u + \overline{d} d|N\rangle \approx 45 \text{~MeV}$ (where $m_l$ denotes the average up and down quark mass). This is entirely consistent with the traditional value for this term determined from $\pi N$ scattering through a dispersion relation analysis~\cite{Brown:1971pn,Gasser:1990ce}. The best value for the strange sigma term, however, has seen an enormous revision over the last two decades. The modern lattice results in Figure~\ref{fig:sigmas} indicate a value for $\sigma_s$ of 20-60~MeV, which is an order of magnitude smaller (and significantly more precise) than the traditional value of this term obtained indirectly using $\sigma_{\pi N}$ and a best-estimate for the singlet contribution $\sigma_0=m_l\langle N| \overline{u}u + \overline{d} d-2\overline{s} s|N \rangle$. This traditional approach yielded values of $\sigma_s$ as large as 300~MeV. Although early lattice studies were compatible with this result, recent work suggests that the values obtained were erroneously large as a result of operator mixing effects~\cite{Bali:2011ks}. Of course, since the strange sigma commutator may be interpreted as the contribution to the mass of the nucleon from the strange quark, a value as large as 300~MeV would indeed be remarkable; it would suggest that almost a third of the nucleon mass arises from non-valence quarks. This appears incompatible with the widely used constituent quark models, for example. Clearly, this issue appears to have been resolved in favor of a smaller strange quark sigma term. Improved precision of the best lattice values of $\sigma_s$ is still extremely desirable, however, particularly in the context of dark matter direct detection experiments~\cite{Ellis2008,Giedt2009,Ellis2012}.

\subsection{The proton-neutron mass difference}
\label{sec:CSV}

Charge symmetry violation (CSV) in the nucleon mass is small---the neutron-proton mass difference is one part in a thousand. The effects of this small CSV, however, are of tremendous significance; it is precisely this which ensures that the hydrogen atom is stable against weak decay and that neutrons can decay into protons (plus electrons and antineutrinos) in radioactive beta decay. 
While the total proton-neutron mass difference is known extremely precisely from experiments~\cite{pdg}, its decomposition into strong and electromagnetic contributions is less well known.
In recent years there has been considerable effort invested in lattice-based determinations of both the QCD contribution to the baryon mass splittings~\cite{Beane2007,Blum2010,WalkerLoud2010,Divitiis2011,Horsley2012,Borsanyi:2013lga} and the electromagnetic contribution~\cite{Duncan1996,Basak2008,Portelli2011,Glaessle2011,Thomas:2014dxa}. However, $1+1+1$--flavour simulations---at this stage the only way to directly probe the full flavor-dependence of QCD observables---are not yet widely available (the first set of $1+1+1+1$--flavour ensembles has recently appeared~\cite{Borsanyi:2014jba}). 
Such studies are of particular interest in the light of recent results which suggest that the accepted value for the electromagnetic contribution to the neutron-proton mass difference calculated using the Cottingham formula may be too small because of an omission in the traditional analysis~\cite{Walker-Loud2012,Leutwyler:2015jga}. 

In this review focused on the ChEFT--lattice-QCD connection we concentrate not on direct lattice calculations of the strong or electromagnetic proton-neutron mass difference (although these also involve EFT to correct for finite-volume effects), but on indirect methods which involve ChEFT input. 
In particular, one can use ChEFT techniques to determine the strong isospin-breaking nucleon mass difference while using as input the high-precision isospin-averaged simulations which are currently available for the octet baryon masses. In a ChEFT expansion of the octet baryon masses (e.g.,~\cite{Lutz:2014oxa}), the unknown low-energy constants are the same whether or not the SU(2) symmetry is broken, that is, whether or not the light quarks are mass-degenerate. Having fit these constants to isospin-averaged simulation results (as one would do to perform a chiral extrapolation), the only additional input needed to deduce the strong proton-neutron mass difference is a value for the light-quark mass ratio $R=m_u/m_d$.
A similar procedure can be followed using a linear and quadratic $\SU(3)$-flavour-symmetry--breaking expansion in the quark masses, provided the average quark mass is kept constant at its physical value (as it is in Ref.~\cite{Horsley2012}). The uncertainties obtained using these indirect methods~\cite{Shanahan:2012wa,Horsley2012} are comparable to those from recent direct lattice simulations~\cite{Borsanyi:2013lga}.

\begin{figure}
\centering
\subfigure[Leutwyler: $R= 0.553(43)$\cite{Leutwyler:1996qg}.]{
\includegraphics[width=0.47\textwidth]{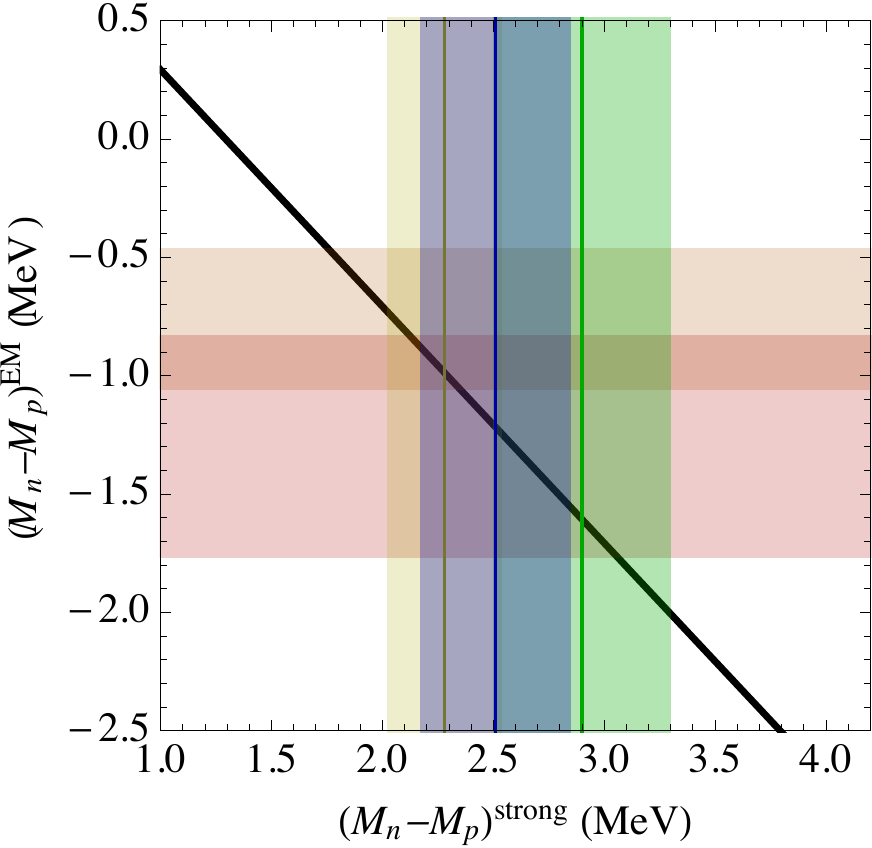}
}
\subfigure[FLAG: $R= 0.47(4)$\cite{FLAG}.]{
\includegraphics[width=0.47\textwidth]{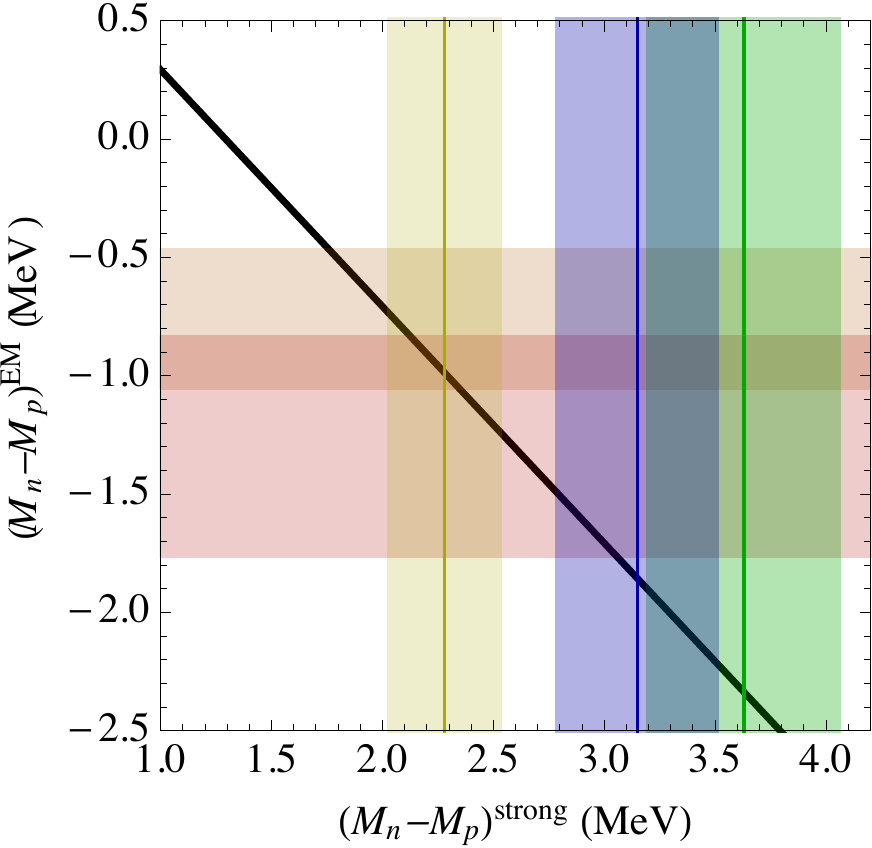}
}
\caption{Strong and electromagnetic contributions to the neutron-proton mass difference. The black line indicates the experimental constraint on the total~\cite{pdg}. The green and blue shaded bands show results from Ref.~\cite{Shanahan:2012wa} of fits to the PACS-CS and QCDSF-UKQCD collaboration simulations, respectively, with the given values of the light-quark mass ratio $R$. The yellow vertical band indicates a direct lattice calculation of the strong mass splitting by the BMW collaboration~\cite{Borsanyi:2013lga} (consistent with \cite{Thomas:2014dxa}). The horizontal bands show the traditional~\cite{Gasser1982} (orange) and Ref.~\cite{Walker-Loud2012} (pink) estimates for the EM contribution.\label{fig:StrongEMBands}}
\end{figure}

Conversely, using this methodology, more precise direct lattice (or phenomenological) determinations of the strong or electromagnetic contributions to the mass splittings may allow a significantly improved determination of $R$. At the current level of precision it is already clear from Fig.~\ref{fig:StrongEMBands} that, for consistency with direct lattice calculations~\cite{Borsanyi:2013lga} and experiment, the analysis of Ref.~\cite{Shanahan:2012wa} using finite-range regulated heavy-baryon ChEFT with decuplet degrees of freedom favors the Leutwyler value $R=0.553(43)$~\cite{Leutwyler:1996qg} over the smaller FLAG result $R=0.47(4)$~\cite{FLAG}. In this way, the relationship between precise isospin-averaged and broken lattice QCD simulations of the octet baryon masses could greatly improve our current best value for the light-quark mass ratio.

\subsection{Charge symmetry violation in the nucleon electromagnetic form factors}

The strange electromagnetic form factors of the nucleon have been the focus of intensive experimental and theoretical effort since the late 1980s when it was realized that they could be determined through measurements of neutral weak current matrix elements by parity-violating electron scattering (PVES)~\cite{Kaplan:1988ku,Mckeown:1989ir,Beck:1989tg}. Since the nucleon has no valence strange quarks these quantities can only arise through quantum fluctuations and hence provide a clean probe of vacuum contributions to nucleon properties. 
At present, the accuracy of theoretical calculations of the strange magnetic moment in particular~\cite{Shanahan:2014tja,Doi:2009sq,Leinweber:2006ug,Leinweber:2004tc} exceeds that of the best experimental values~\cite{Young:2007zs} by almost an order of magnitude, which is a remarkable exception in strong-interaction physics.
A significant limiting factor in future experimental determinations of the strange form factors through PVES experiments at Mainz~\cite{Maas:2004dh,Baunack:2009gy} and JLab~\cite{Aniol:2005zg,Aniol:2005zf,Acha:2006my} is theoretical, arising from the assumption of good charge symmetry in the electromagnetic form factors of the nucleon. 

\begin{figure}
\centering
\subfigure[]{\label{fig:CSVFFMTheirs}
\includegraphics[width=0.475\textwidth]{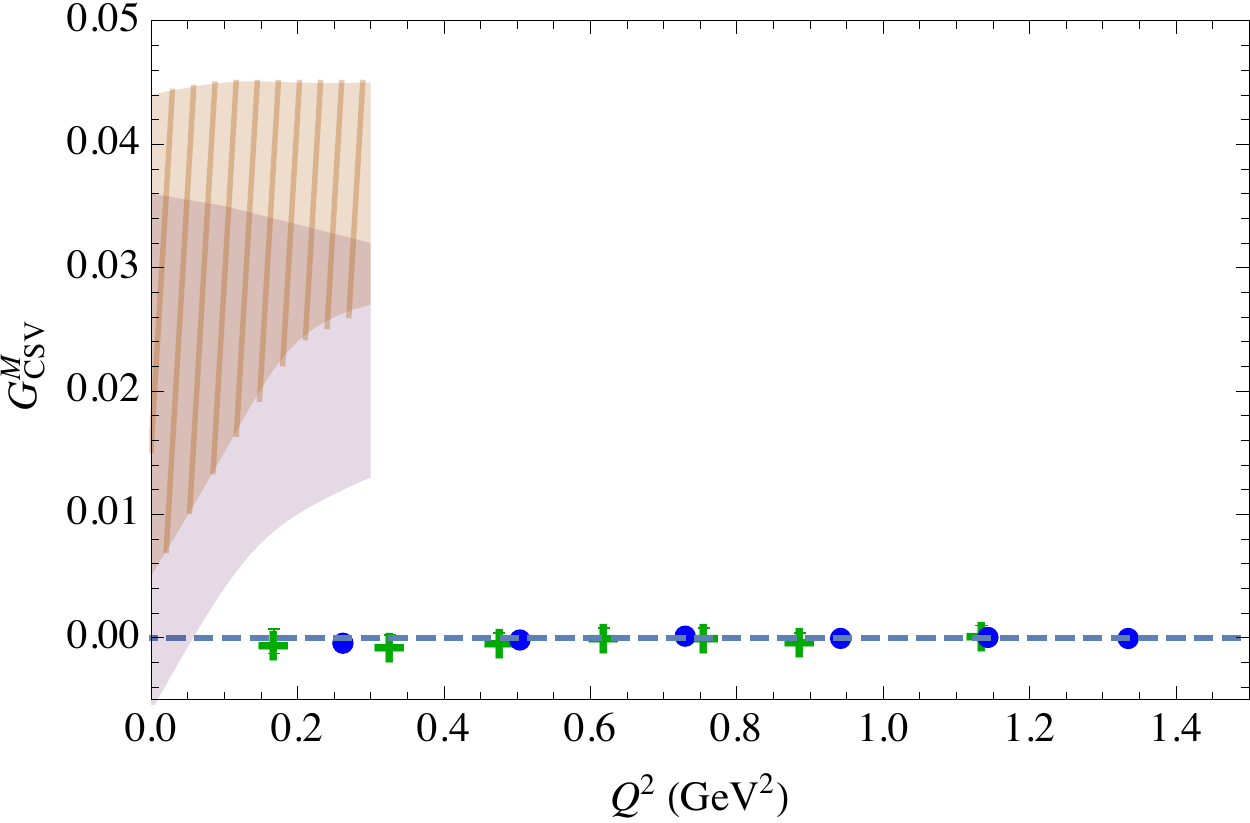}
}
\subfigure[]{\label{fig:CSVFFETheirs}
\includegraphics[width=0.475\textwidth]{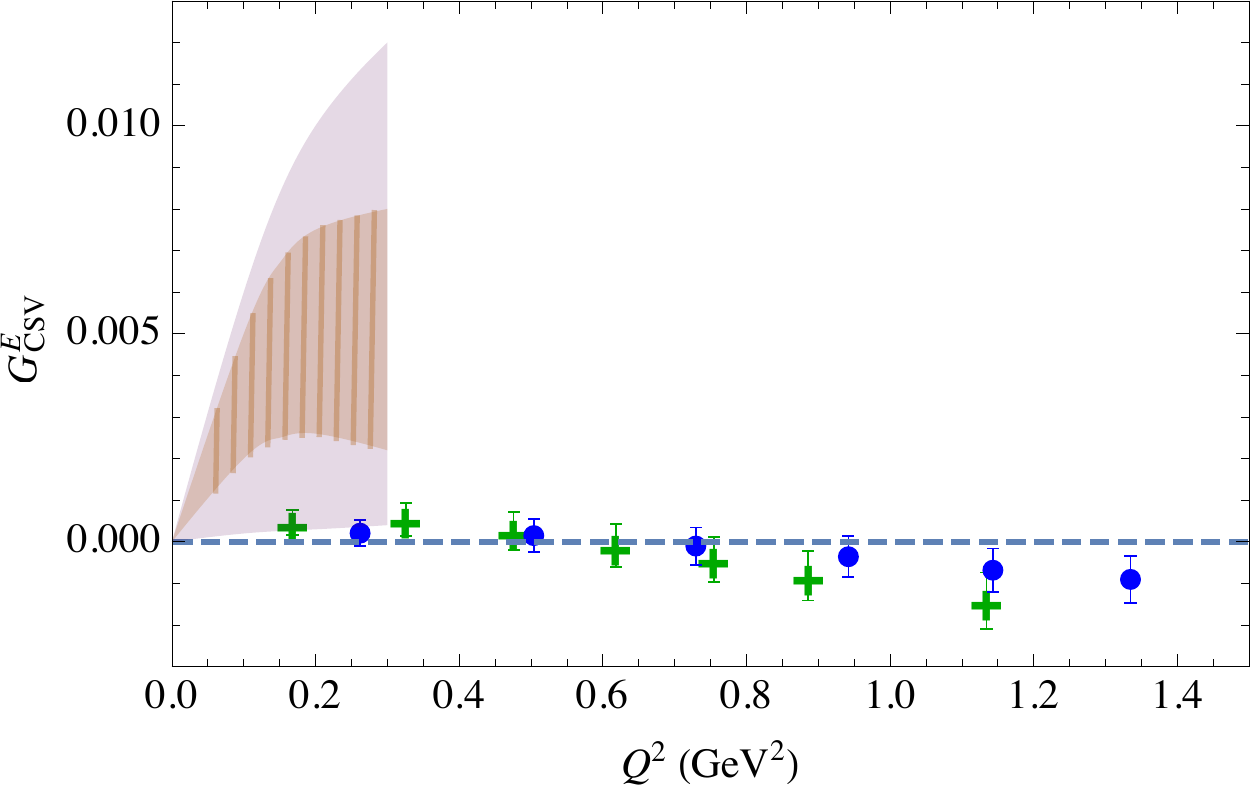}
}
\caption{\label{fig:TotalCSVTheirs}Figure from Ref.~\cite{Shanahan:2015caa} showing lattice QCD / ChEFT results for the magnetic and electric CSV form factors as relevant to experimental determinations of nucleon strangeness as blue and green points. The orange (striped) and purple (plain) bands show model calculations of these terms taken from figures in Refs.~\cite{Kubis:2006cy} and \cite{Wagman:2014nfa}, respectively (where in the latter case the bands shown span the full range of results given for various choices of the model parameters).}
\end{figure}

Since theoretical predictions of the size of the charge symmetry violating (CSV) form factor $G_\textrm{CSV}$ vary through several orders of magnitude~\cite{Wagman:2014nfa,Kubis:2006cy,Miller:2006tv}, lattice determinations of this quantity are of particular interest and relevance. At this stage, however, there are no isospin-broken ($N_f=1+1+1$) simulations of the electromagnetic form factors available. However, the same procedure described in Section~\ref{sec:CSV} above for the determination of isospin-breaking results from isospin-averaged lattice simulations has recently been applied to this quantity using a finite-range regulated heavy-baryon ChEFT formalism~\cite{Shanahan:2015caa}.
These results, shown in shown in Fig.~\ref{fig:TotalCSVTheirs} compared to the previous best theory determinations of $G_\textrm{CSV}$, give quantitative confirmation that CSV effects in the electromagnetic form factors, for momentum transfers up to approximately $1.3\GeV^2$, are at the level of 0.2\% of the relevant proton form factors---an order of magnitude smaller than the precision of existing parity-violating electron scattering studies. Independent confirmation of these significant results, either through isospin-broken lattice simulations directly or through a similar analysis, would be very valuable.

\section{Concluding remarks}

For many years ChEFT techniques have supported lattice QCD calculations of hadron observables by allowing systematic effects from unphysical lattice parameters to be controlled. In the era of precision lattice simulations approaching the physical point these methods remain important. This very brief review has only touched on the applications of ChEFT to modern lattice QCD. 

After a brief discussion of the most widely-used ChEFT formalisms in this context, we enumerated some examples of recent simulations where chiral, volume, and/or momentum extrapolation was essential to achieving physically-relevant results. For example, we considered simulations of the nucleon electromagnetic form factors, which are of particular interest in the light of the proton radius puzzle, and approaches to determining the hadronic vacuum polarization contribution to muon $g-2$. Even as these and future simulations approach the physical point, ChPT is still of use; by varying the quark mass away from the physical value, one might discover that the numerical results do not match the predictions of continuum ChPT, for instance because the lattice spacing is still too large, the volume too small, or because of some other systematic effect. In other words, ChPT, assuming convergence, provides a useful tool for the validation of results obtained with lattice QCD. 
For some observables, direct physical-mass lattice simulations are still unachievable computationally and ChEFT methods are integral to the extraction of physical results now and will be for some time into the foreseeable future. One example discussed in this review is the nucleon polarizabilities, for which numerical simulations are only now becoming available. Another example is the electromagnetic decays of the antitriplet and sextet charmed baryon systems, for which lattice simulation results are now available~\cite{Brown:2014ena} and chiral extrapolation techniques have recently been studied~\cite{Jiang:2015xqa}. Moving beyond hadron structure to preliminary studies of light nuclei these techniques are of course all the more important; only a few properties of nuclei have so far been calculated near the physical point and higher precision is required in all of these calculations in order to impact experimental programs~\cite{Beane:2014oea}.

As well as facilitating chiral, momentum and volume extrapolations, ChEFT techniques also allow access to observables other than those directly simulated through the determination of universal low-energy constants. In this way, for example, isospin-averaged simulations can give information on isospin-breaking quantities, and omitted disconnected loops can be restored in partially-quenched simulations. Often these approaches can give information about physically interesting observables long before they are directly accessible to the lattice. We discussed in particular determinations of charge-symmetry violation in the nucleon masses and electromagnetic form factors, which capitalize on the high-precision $2+1$-flavor lattice simulations currently available for these quantities.

Clearly, ChEFT techniques have an important role to play now and in the future of lattice QCD. As higher precision is reached and new observables---both hadronic and nuclear---are calculated through lattice techniques for the first time, ChEFT techniques will provide an invaluable guide to the systematic effects naturally associated with the lattice formalism. Moreover, these techniques provide interpretations of simulation results based on low-energy QCD as well as experiment-based predictions, given experimental determinations of the LECS, of observables that have not or cannot be measured explicitly. In this way ChEFT is a natural companion to the lattice approach.

\section*{Acknowledgments}

This work is supported by the U.S. Department of Energy under Grant Contract Number DE-SC00012567.

\section*{References}
\bibliography{ReviewBib}

\end{document}